\documentclass[5p,twocolumn]{elsarticle}
\usepackage[USenglish]{babel}
\usepackage[T1]{fontenc}
\usepackage[utf8]{inputenc}
\usepackage{amssymb} 
\usepackage{amsmath}
\usepackage{amsthm}
\usepackage{amsfonts}
\usepackage{siunitx}
\usepackage{booktabs}

\journal{Nuclear Instruments and Methods in Physics Research Section A}

\begin{document}


\title{Simulations and plans for possible DLA experiments at SINBAD}

\author[desy,uhh]{F.~Mayet}
\ead{frank.mayet@desy.de}

\author[desy]{R.~Assmann}
\author[desy]{J.~Boedewadt}
\author[desy]{R.~Brinkmann}
\author[desy]{U.~Dorda}
\author[desy,uhh]{W.~Kuropka}
\author[desy]{C.~Lechner}
\author[desy]{B.~Marchetti}
\author[desy,uhh]{J.~Zhu}

\address[desy]{Deutsches Elektronen-Synchrotron DESY,
Notkestraße 85, 22607 Hamburg, Germany}
\address[uhh]{Universität Hamburg, Institut für Experimentalphysik,
Luruper Chaussee 149, 22761 Hamburg, Germany}

\begin{abstract}
In this work we present the outlines of possible experiments for dielectric laser acceleration (DLA) of ultra-short relativistic electron bunches produced by the ARES linac, currently under construction at the SINBAD facility (DESY Hamburg). The experiments are to be performed as part of the Accelerator on a Chip International Program (ACHIP), funded by the Gordon and Betty Moore Foundation. At SINBAD we plan to test the acceleration of already pre-accelerated relativistic electron bunches in laser-illuminated dielectric grating structures. We present outlines of both the acceleration of ultra-short single bunches, as well as the option to accelerate phase-synchronous sub-fs microbunch trains. Here the electron bunch is conditioned prior to the injection by interaction with an external laser field in an undulator. This generates a sinusoidal energy modulation that is transformed into periodic microbunches in a subsequent chicane. The phase synchronization is achieved by driving both the modulation process and the DLA with the same laser pulse. In addition to the conceptual layouts and plans of the experiments we present start-to-end simulation results for different ARES working points.
\end{abstract}


\maketitle



\section{Introduction}
The Accelerator on a Chip International Program (ACHIP) funded by the Gordon and Betty Moore Foundation aims to demonstrate a working prototype of a particle accelerator on a chip until 2021. Being part of the ACHIP collaboration DESY will conduct related test experiments at its SINBAD facility \citep{Dorda:IPAC2017-MOPVA012}. Here we present plans for the first DLA experiments at SINBAD using electrons produced by the ARES linac \citep{Marchetti:IPAC2017-TUPAB040,Zhu:2016fh}. In addition to the conceptual layout of the experiment we present possible linac working points and an estimation of the expected results using a $\beta$-matched dual grating type DLA structure~\cite{Peralta:2013vpa} illuminated by a 1 or 2 micron laser. 

At the time of this publication three different ACHIP-related experiments are planned to be conducted at ARES. 

In the first experiment we plan to accelerate already pre-accelerated relativistic single electron bunches in a laser-illuminated dielectric grating structure. The goal of the experiment is to show net-acceleration with low energy spread growth instead of a broad energy modulation. Since the ACHIP target periodicity of the DLA structure currently is \SI{2}{\micro\meter}, achieving a DLA drive laser to electron phase spread of $\sigma _\phi < \pi/4$ is challenging. At ARES we aim to produce sufficiently short bunches using the velocity bunching technique \cite{MARCHETTI2016278,Serafini200186,PhysRevLett.104.054801,Mayet:IPAC2017-WEPVA007}.

In addition to the short bunch lengths the small scale of the DLA structures consequentely implies the need for ultra-stable drive laser to electron phase. In order to enable stable multi-shot operation we plan to also implement a technique for phase-synchronous acceleration of microbunch trains \cite{Mayet:IPAC2017-WEPVA006}. The main idea of the scheme is to condition a relatively long bunch in a way that it is transformed into a train of ultra-short microbunches. If done correctly, these microbunches then populate the periodic accelerating buckets in a phase stable manner.

The third experiment, which is planned to be conducted at ARES, aims to test the feasibility of using DLA structures as transverse deflectors (DLA TDS). To this end either single bunches or trains of microbunches are injected in a suitably designed and driven structure. A detailed study of this topic can be found in these proceedings \citep{WILLI_TDS}. Hence it is left out here for brevity.

\section{The SINBAD Facility and ARES}
\begin{figure}[!htb]
   \centering
   \includegraphics*[width=0.47\textwidth]{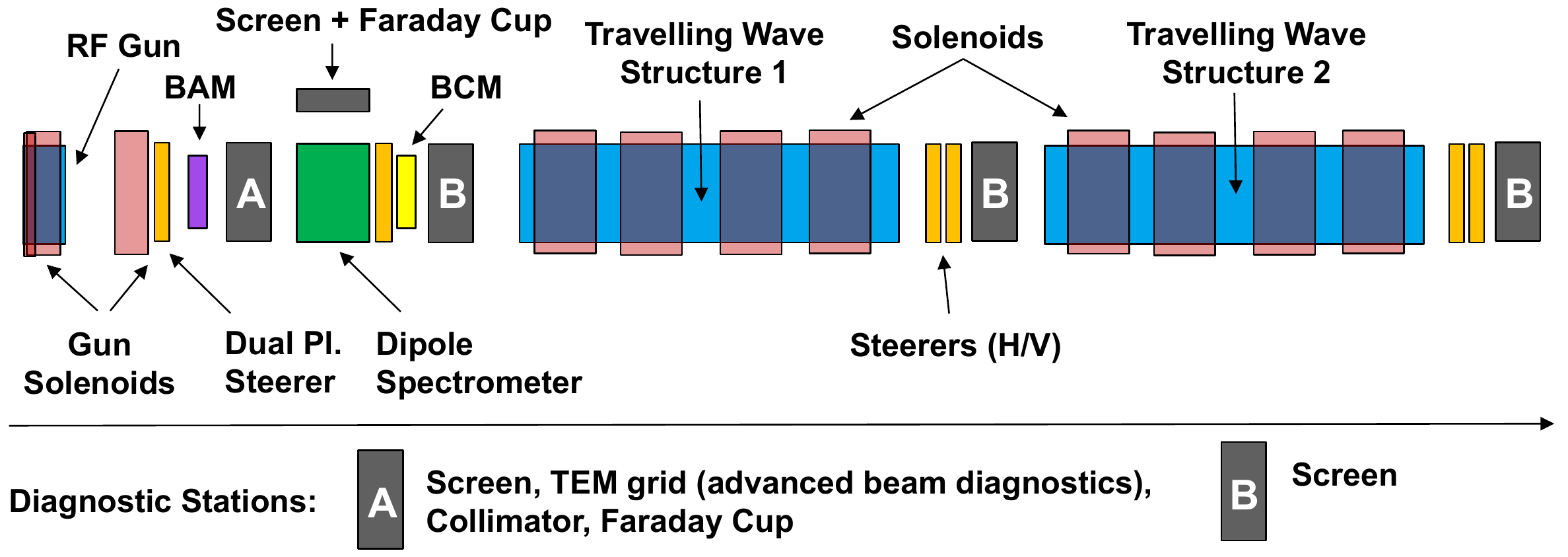}
   \caption{The ARES linac.}
   \label{fig:ARES_linac}
\end{figure}

The dedicated accelerator R\&D facility SINBAD (Short INnovative Bunches and Accelerators at Desy) is a new facility at DESY, which is foreseen to host multiple independent experiments. Two experiments are currently under construction: ARES (Accelerator Research Experiment at Sinbad) and AXSIS \citep{AXSIS}. The ACHIP experiments are planned to be conducted at the ARES linac (see Fig.~\ref{fig:ARES_linac}). 
The ARES linac is a conventional 100\,MeV S-band linac, which is designed to produce sub-fs electron bunches with charges in the range of 0.5-20\,pC. To this end the electrons are first accelerated in a 1.5 cell S-band gun to 5\,MeV with a single bunch repetition rate of up to 50\,Hz. They are then further accelerated in two S-band traveling wave (TW) structures. In order to compress the bunch the two TW structures can be used for velocity bunching \citep{MARCHETTI2016278}. Transverse focusing is achieved using multiple solenoids (cf. Fig. \ref{fig:ARES_linac}). The accelerator is planned to be available for DLA-related experiments in mid 2019.

\begin{figure}[!tbh]
   \centering
   \includegraphics*[width=0.47\textwidth]{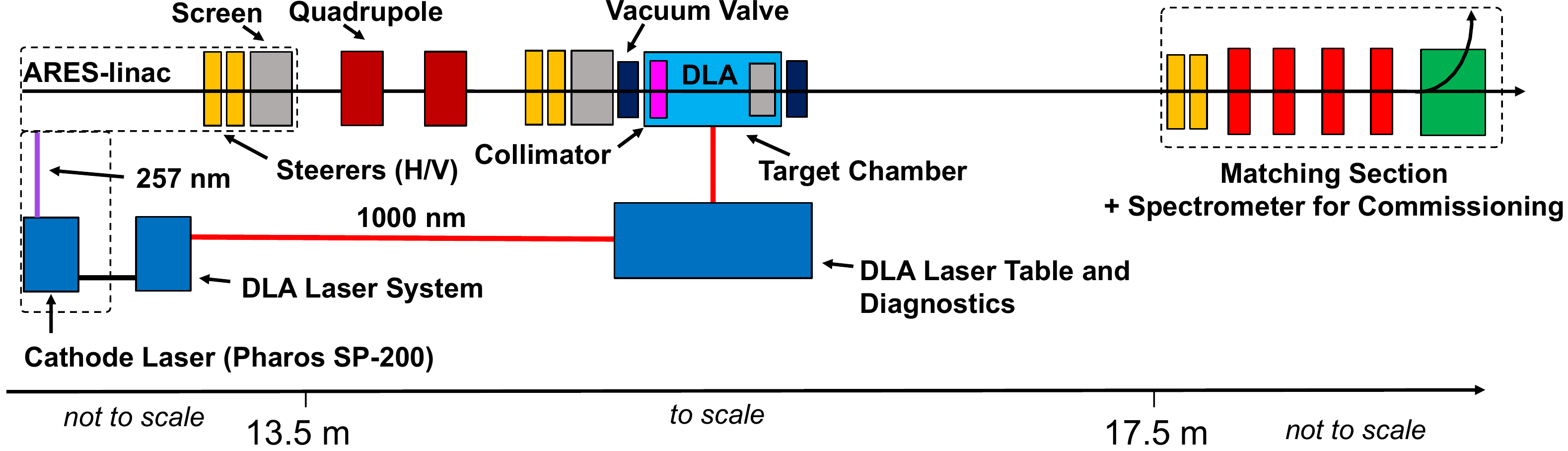}
   \caption{Sketch of the ARES beam line and the preliminary layout of the first ACHIP experiment.}
   \label{fig:Stage0}
\end{figure}

\begin{figure}[!tbh]
   \centering
   \includegraphics*[width=0.47\textwidth]{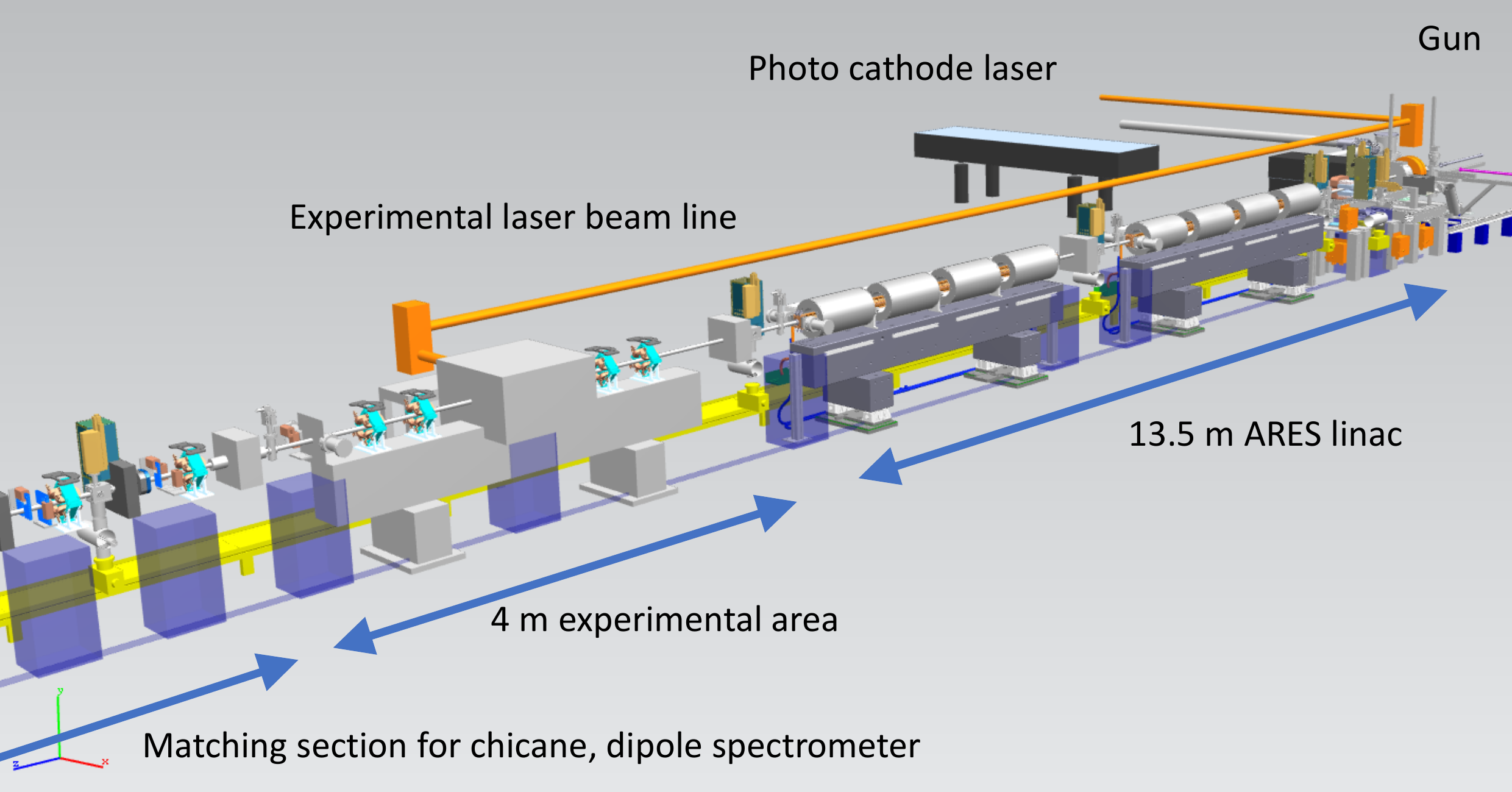}
   \caption{CAD model of the relevant part of the ARES linac beamline and the experimental area, which is foreseen for the ACHIP-related experiments.}
   \label{fig:Stage0CAD}
\end{figure}

In the first stage the ACHIP experiments are planned to be conducted within the 4\,m space between the two TW structures and a matching region, which is needed for a later energy upgrade of the beamline (see Fig.~\ref{fig:Stage0} and Fig.~\ref{fig:Stage0CAD}). After the interaction point the beam will be transported through the matching region to a dipole spectrometer in order to diagnose the energy gain/modulation.
\section{External Injection of Single Bunches}
The first DLA-related experiments at SINBAD are planned to be based on the external injection of single ultra-short electron bunches into a grating-type DLA structure. In the current design both the cathode of the RF-gun and the DLA are driven by the same laser system (see Fig.~\ref{fig:Stage0}). Since the DLA is foreseen to be operated with a laser wavelength of \SI{2}{\micro\meter}, the initial \SI{1028}{\nano\meter} beam is split and converted on the one hand to \SI{257}{\nano\meter} (fourth harmonic generation) and on the other to \SI{2}{\micro\meter} (optical parametric amplifier). This setup has the advantage of intrinsic synchronization between the cathode and the DLA laser beams. The relative electron to laser phase jitter is hence mainly given by the RF-induced beam arrival time jitter contribution.
\subsection{Working Point}
Tab.~\ref{tab:WP0} shows the beam parameters at the interaction point for a simulated ARES working point. It is based on a \SI{100}{\femto\second} rms laser pulse length on the cathode (Gaussian time profile) and optimized for minimal bunch length using the velocity bunching technique. The simulation was performed using ASTRA \citep{ASTRA} including space charge.
\begin{table}[hbt]
   \centering
   \caption{Simulated Working Point for External Injection of Single Bunches.}
   \vspace{0.2cm}
   \begin{tabular}{ll}
      \toprule
      \textbf{Parameter @ IP} & \textbf{Value}\\
      \midrule
      Charge [pC] & 0.5\\
      Bunch Length [fs, FWHM] & 2.1\\
      E [MeV] & 99.1\\
      $\Delta$E/E [\%] & 0.12\\
      $\sigma _{xy}$ [$\mu$m] & 7.8\\
      $\epsilon _{n,xy}$ [nm] & 105\\
      \bottomrule
   \end{tabular}
   \label{tab:WP0}
\end{table}
\subsection{Simulation}
\begin{figure}[!tbh]
   \centering
   \includegraphics*[width=0.47\textwidth]{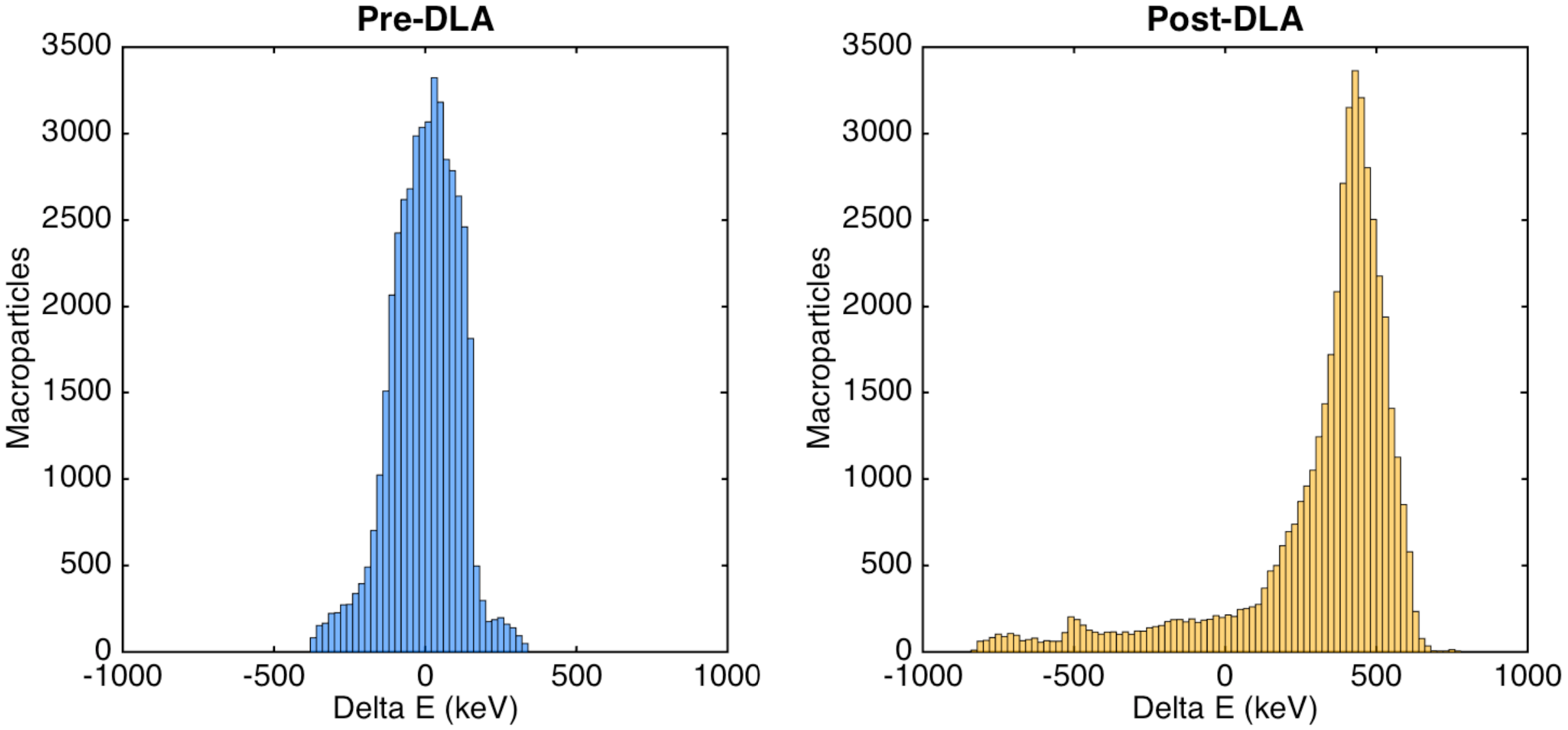}
   \caption{Simulated energy spectrum of the transmitted part of the bunch before (left) and after (right) the DLA interaction (mean energy gain of $\sim 300$\,keV). The bins are chosen according to the expected minimum energy resolution of the spectrometer ($\sim10^{-4}$).}
   \label{fig:energyspectrum}
\end{figure}

Simulations were performed using a combination of ASTRA and VSim 7.2 \citep{VSim}. The procedure combines the simulation of the ARES working point up to the DLA (ASTRA-based) and the DLA interaction (VSim-based). Any possible interaction of the electrons with the dielectric material is currently not taken into account. The beam is assumed to be collimated just upstream of the DLA. Figure \ref{fig:energyspectrum} shows the energy spectrum of the transmitted part before and after the DLA interaction respectively. In the simulation an accelerating gradient of 1\,GeV/m over 150 periods was assumed. In reality this will depend a lot on the achievable laser parameters at the DLA, which are not fixed at the time of this publication. As the electrons are already highly relativistic dephasing is not an issue here. The results show that $\sim$80\,\% of the collimated core of the bunch is accelerated.
\section{Phase-Synchronous Acceleration} \label{sec:phasesyncacc}
In the previous section a first possible experiment involving single bunches was described. As has been stated above achieving an energy spectrum as shown in Fig.~\ref{fig:energyspectrum} over many conscutive shots assumes a very stable bunch arrival time at the DLA. In order to achieve reasonably low energy spread growth due to arrival time jitter, the rms phase stability needs to be $< \pi /4$, which translates to $< 1$\,fs for a DLA period length of \SI{2}{\micro\meter}. This is a very challenging goal, as the design phase stability of ARES is currently given as $< 10$\,fs rms.

In order to tackle this problem, we want to adapt a scheme that has already been successfully used in other contexts \citep{Sears,Kimura:2001ik} to our DLA case. The main idea of the scheme is to condition a relatively long bunch in a way that it is transformed into a train of ultra- short microbunches. If done correctly, these microbunches then populate the periodic accelerating buckets in a phase stable manner.
\begin{figure}[!htb]
   \centering
   \includegraphics*[width=0.47\textwidth]{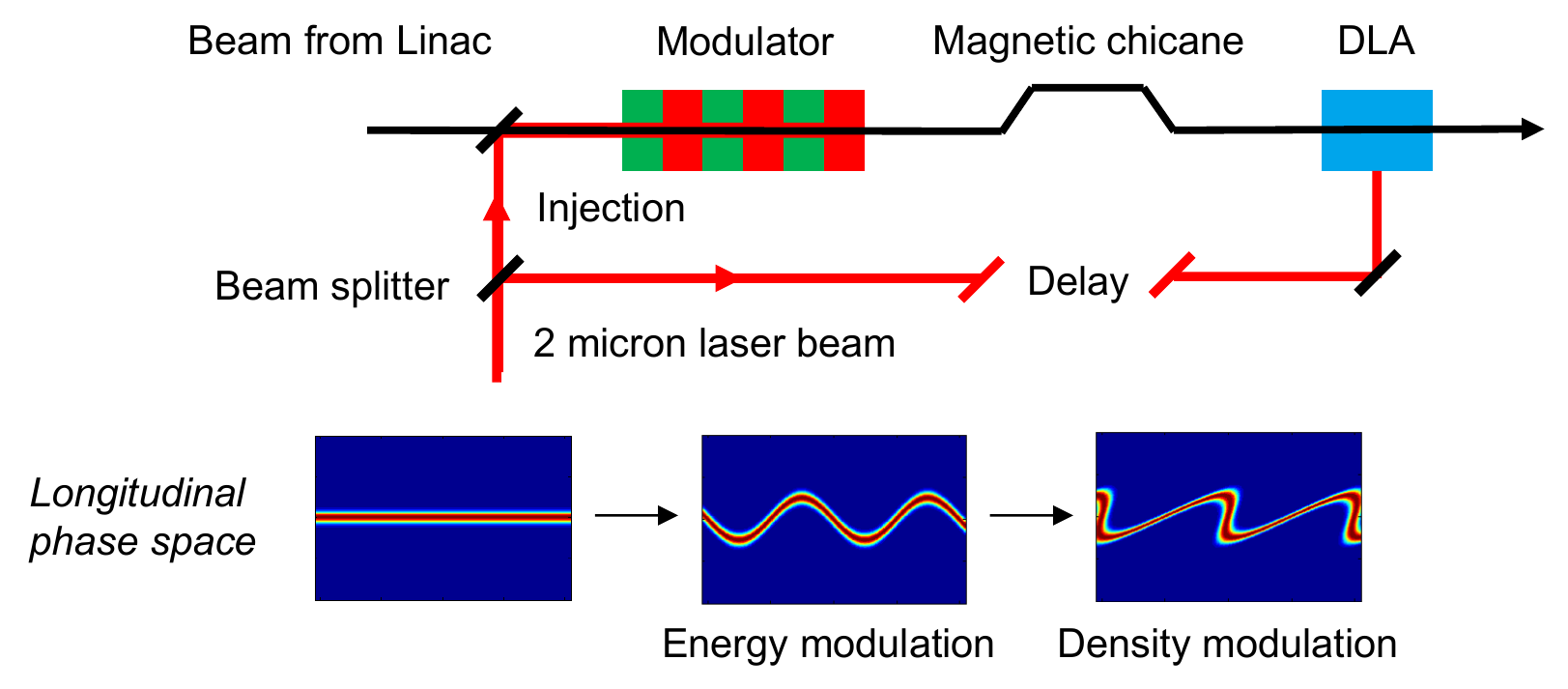}
   \caption{Basic representation of the microbunching scheme using a modulator and a chicane.}
   \label{fig:Setup}
\end{figure}
Fig.~\ref{fig:Setup} shows a sketch of how the scheme works. The incoming long electron bunch is modulated inside an undulator, into which a laser pulse is colinearly injected in a way that it overlaps with the electron bunch in time and space. The external laser field consequently imprints an energy modulation onto the electron distribution, which can in the plane wave approximation be expressed as \cite{Hemsing:2014ju}
\begin{equation}
  \Delta \gamma = \sqrt{\frac{P_L}{P_0}}\frac{2KL_u\mathcal{J}}{\gamma w_0} \cos (k_L s),
  \label{eq:modulation}
\end{equation}
where $s$ is the comoving longitudinal coordinate, $P_L$ the laser beam power, $P_0 \approx 8.7$\,GW, $K$ the undulator parameter, $L_u$ the undulator length, $w_0$ the laser waist, $k_L$ the laser wave number and $\mathcal{J} = J_0(\xi/2)-J_1(\xi/2)$ with $\xi = K^2/(2+K^2)$. The device is hence called a \emph{laser modulator}. The imprinted energy modulation can then be transformed into a density modulation, which ultimately (and ideally) results in a train of microbunches. If the energy of the incoming electrons is low enough this transformation can be achieved by a simple drift, but in case of highly relativistic electrons a dispersive section such as a magnetic chicane has to be used. Once the train of microbunches is formed it can be transported and injected into the DLA structure. If both the modulator and the DLA are driven by the same laser and the relative phase jitter between the two arms is negligible, intrinsic phase synchronisation between the microbunches and the DLA field can be achieved. Any laser to electron bunch phase jitter in the DLA caused by the laser system or the electron time of arrival is compensated due to the fact that the intrabunch phase of the microbunch train is also shifted by the same amount. In other words: The modulator acts as a focusing device in the time/phase domain. 
Any residual laser to electron phase jitter can now be attributed to the mean electron energy (via $R_{56}$) and the phase jitter between the two laser arms.

\subsection{Integration}
In \citep{Mayet:IPAC2017-WEPVA006} we have already studied a theoretical extension of the ARES beamline in order to show the feasibility of the concept in the \SI{2}{\micro\meter} DLA context. A clear enhancement of the accelerated fraction in the modulator on case was confirmed in simulation compared to the non-modulated beam. Preliminary start to end simulations based on ASTRA, GENESIS 1.3 \citep{GENESIS} and VSim 7.2 showed an achievable microbunch train with microbunches as short as $(699 \pm 88)$\,attoseconds\,FWHM with a spacing of $(2.00 \pm 0.01)$\,\SI{}{\micro\meter} in its 6 period core.

In order to keep the setup at SINBAD as small and affordable as possible we aim to use a scaled down version of the setup, which fits into the planned target chamber. Current plans foresee repurposing components that have already been used at SLAC (NLCTA) \citep{Sears}. The components comprise a miniature undulator, permanent magnetic chicance and permanent magnetic quadrupole (PMQ) triplet. Tab.~\ref{tab:modhardware} shows the relevant parameters of the components.
\begin{table}[hbt]
   \centering
   \caption{Relevant Parameters of the Microbunching Hardware according to \citep{Sears}. Note that all Components can be adjusted.}
   \vspace{0.2cm}
   \begin{tabular}{ll}
      \toprule
      \textbf{Parameter} & \textbf{Value}\\
      \midrule
      Undulator & \\
      \hspace{0.35cm}Undulator Period [cm] & 1.8\\
      \hspace{0.35cm}Periods & 3\\
      \hspace{0.35cm}Undulator Parameter & 0.46-1.7\\
      Chicane & \\
      \hspace{0.35cm}$R_{56}$ [mm] & 0.06-0.22\\
      PMQ & \\
      \hspace{0.35cm}Bore Radius [mm] & 3\\
      \hspace{0.35cm}Magnet Length (1) [mm] & 7\\
      \hspace{0.35cm}Magnet Length (2,3) [mm] & 13\\
      \hspace{0.35cm}Spacing (adjustable) [mm] & 4-20\\
      \hspace{0.35cm}$B_\text{max,pole}$ [T] & 0.6\\      
      \bottomrule
   \end{tabular}
   \label{tab:modhardware}
\end{table}
\subsection{Simulation}
\begin{figure}[!htb]
	\centering
	\includegraphics*[width=0.47\textwidth]{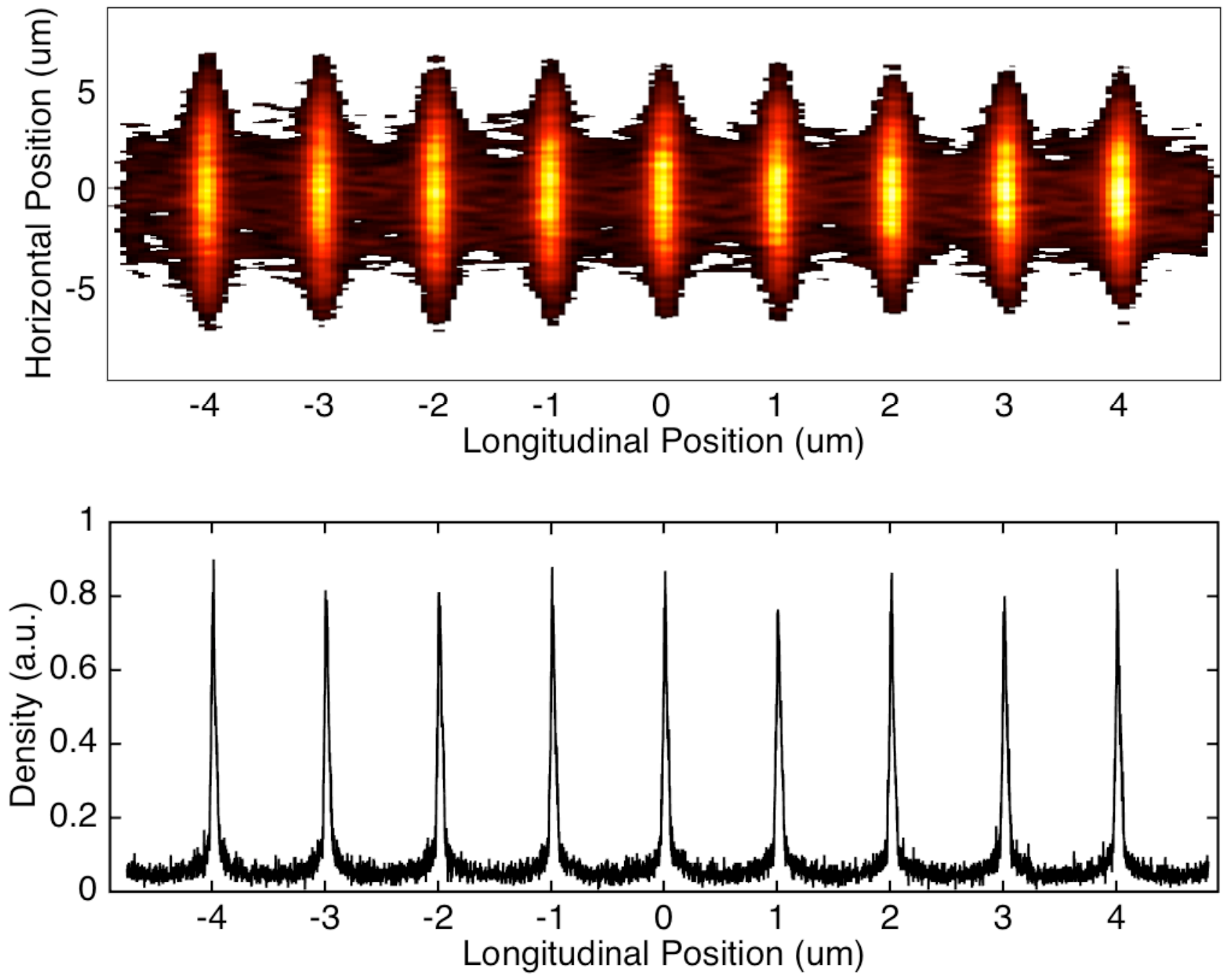}
	\caption{\textbf{Top:} $z-x$ phase space of a longitudinal 9 period slice of the simulated beam at the DLA entrance showing 9 consecutive microbunches. \textbf{Bottom:} Projection across the horizontal axis. Microbunch length: $(160 \pm 13)$\,as\,FWHM, spacing: $(1.0011 \pm 0.0068)$\,\SI{}{\micro\meter}.} 
	\label{fig:microbunches}
\end{figure}

\begin{figure}[!htb]
	\centering
	\includegraphics*[width=0.47\textwidth]{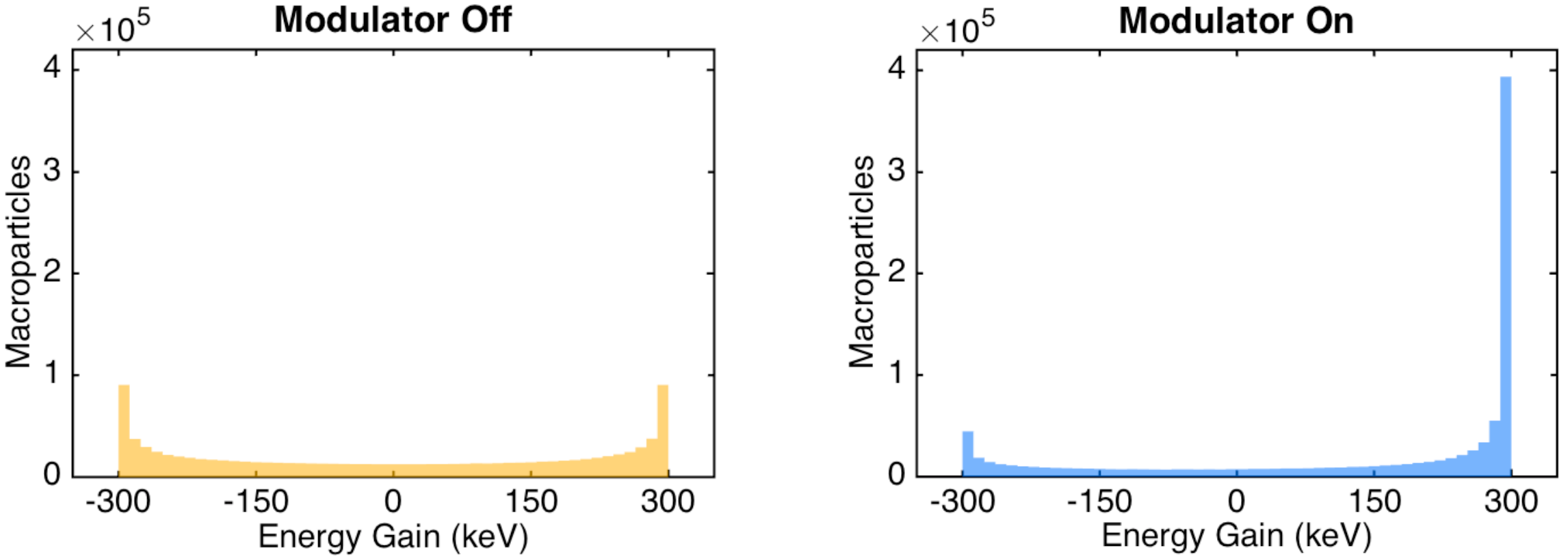}
	\caption{Simulated energy gain spectrum of the transmitted/collimated part of the microbunched beam after the DLA interaction. Acceleration is enhanced in the modulated case.}
	\label{fig:gainspectrum}
\end{figure}

The undulator was originally designed for an \SI{800}{\nano\meter} drive laser and 60\,MeV. The wavelength of emitted undulator radiation is given by
\begin{equation}
	\lambda _l = \frac{\lambda _u}{2 \gamma ^2} \left( 1 + \frac{K^2}{2}\right),
	\label{eq:undulatorrad}
\end{equation}
where $\lambda _u$ is the undulator period and $\gamma$ the normalized energy of the electrons. Since the pulse energy of our cathode laser system is limited and the conversion from \SI{1}{\micro\meter} to \SI{2}{\micro\meter} involves substantial losses (efficiency $\sim 0.2$), we currently plan to perform the microbunching at \SI{1}{\micro\meter}. Using Eq.~\ref{eq:undulatorrad} hence yields a resonant energy between 50.5 and 75.5\,MeV depending on the adjustable $K$ of the undulator. In order to achieve as short microbunches as possible the relative energy spread at the modulator needs to be as low as possible \citep{Mayet:IPAC2017-WEPVA006}. In our case in order to achieve $\leq \pi/4$ bunch length, $\Delta \gamma / \gamma _0 \leq 5.7 \cdot 10^{-4}$. Also the energy chirp should be as low as possible in order ensure even spacing between the individual microbunches. Tab.~\ref{tab:WPMB} shows a possible ARES working point.
\begin{table}[hbt]
   \centering
   \caption{Simulated Working Point for External Injection of Microbunch Trains.}
   \vspace{0.2cm}
   \begin{tabular}{ll}
      \toprule
      \textbf{Parameter @ Mod} & \textbf{Value}\\
      \midrule
      Charge [pC] & 2.5\\
      Bunch Length [fs, rms] & 112.3\\
      E [MeV] & 75.26\\
      $\delta$ & $0.76 \cdot 10^{-4}$\\
      $\epsilon _{n,xy}$ [nm] & 248\\
      \bottomrule
   \end{tabular}
   \label{tab:WPMB}
\end{table}
For this study we use ELEGANT \citep{ELEGANT} to simulate the laser modulator and OCELOT \citep{Tomin:IPAC2017-WEPAB031} for the further beam transport through the chicane and the PMQ triplet. NB: No collective effects are taken into account in the laser modulator. The OCELOT simulation includes both space charge and CSR. Fig~\ref{fig:microbunches} shows a slice of the $z-x$ phase space of the simulated beam at the DLA entrance, as well as the obtained microbunch properties. The microbunch length along the whole macro pulse is $(186 \pm 73)$\,as\,FWHM with a spacing of $(1.0002 \pm 0.0091)$\,\SI{}{\micro\meter} (N = 118). This is well below our $\sigma _\phi < \pi/4$ requirement at \SI{1}{\micro\meter}. Prior to the DLA simulation the beam is collimated in order to accomodate the <\SI{2}{\micro\meter} channel width.
\begin{table}[hbt]
   \centering
   \caption{Simulation parameters used in ELEGANT and OCELOT.}
   \vspace{0.2cm}
   \begin{tabular}{ll}
      \toprule
      \textbf{Parameter} & \textbf{Value}\\
      \midrule
      Undulator & \\
      \hspace{0.35cm}Undulator Parameter & 1.7\\
      Dispersive Section & \\
      \hspace{0.35cm}$R_{56}$ [mm] & 0.13\\
      Laser & \\
      \hspace{0.35cm}Pulse Energy [\SI{}{\micro\joule}] & 90\\
      \hspace{0.35cm}Pulse Length [fs, rms] & 300\\
      \hspace{0.35cm}Waist [\SI{}{\micro\meter}] & 500\\   
      \bottomrule
   \end{tabular}
   \label{tab:simparams}
\end{table}
Fig.~\ref{fig:gainspectrum} shows the achieved energy gain of the slice due to the DLA interaction for the microbunched and unmodulated beam for comparison. As for the single bunch case we simulated a 150 period dual grating DLA with is operated at an accelerating gradient of 1\,GeV/m. It can be seen that the acceleration efficiency is clearly enhanced as can also be seen in our previous studies \citep{Mayet:IPAC2017-WEPVA006}.
\section{Conclusion}
We have presented two experiments, which are planned to be conducted at the SINBAD facility using electrons produced by the ARES linac. The current schedule foresees first tests using single bunches in mid 2019. As a second stage we aim to show phase-synchronous acceleration of microbunch trains using the setup presented in Sec.~\ref{sec:phasesyncacc}. This setup can potentially enable more efficient and stable acceleration, as well as higher accelerated charges compared to previous experiments. In addition to that the concept of a DLA-based TDS will be further explored \citep{WILLI_TDS}.
\section{Acknowledgments}
This research is funded by the Gordon and Betty Moore Foundation as part of the Accelerator on a Chip International Program (GBMF4744). We would like to thank our ACHIP colleagues J.~England, and K.Wootton for fruitful discussions and support. We also thank our colleages I.~Hartl, L.~Winkelmann and S.~Pumpe from the DESY laser group for their support.

\section*{References}

\bibliographystyle{elsarticle-num}
\bibliography{\jobname}

\end{document}